\documentclass[10pt,twocolumn,letterpaper]{article}

\usepackage[T1]{fontenc}
\usepackage[pagenumbers]{cvpr}%

\usepackage{multirow}
\usepackage{booktabs}
\usepackage{tikz}
\usetikzlibrary{positioning,arrows.meta,calc}

\usepackage[pagebackref,breaklinks,colorlinks]{hyperref}
\providecommand{\etal}{\emph{et al}\onedot}

\begin{document}
\raggedbottom%

\title{Deep-Learning Algorithm for Indeterminate Thyroid Nodules:\\Comparison and Combination with Molecular Testing}

\author{
Jichen Yang$^{1}$\thanks{Corresponding author: \texttt{jy168@duke.edu}.} \quad
Jeffrey Ashton$^{2}$ \quad
Jikai Zhang$^{1}$\\
Maciej A.\ Mazurowski$^{1,2}$ \quad
Benjamin Wildman-Tobriner$^{2}$\\
$^{1}$Department of Electrical \& Computer Engineering, Duke University, Durham, NC, USA\\
$^{2}$Department of Radiology, Duke University Medical Center, Durham, NC, USA\\
}
\maketitle

\begin{abstract}
Thyroid nodules with indeterminate cytology (Bethesda III) carry an intermediate risk of malignancy, and molecular testing is widely used to decide between surgery and surveillance, but it is costly and not universally available. We investigated whether a convolutional neural network (CNN) applied to routine ultrasound images can stratify Bethesda III nodules, and how it compares and combines with molecular testing. We retrospectively identified 466 nodule records from 418 patients who underwent molecular testing (Afirma or ThyroSeq). A CNN pre-trained on determinate nodules was adapted to indeterminate nodules with soft labels and transfer learning. After removal of duplicate records, 119 nodules with surgical pathology were used for development and 47 were held out for testing. On the test set, the deep learning (DL) model achieved an AUC of 0.73 (95\% CI 0.51--0.94), compared with 0.55 (0.38--0.71) for molecular testing (P = 0.15). At a cut-off inherited from the original algorithm, the DL model was more specific than molecular testing (0.95 vs 0.32; P $<$ 0.001), with a higher positive predictive value (0.67 vs 0.21), a lower sensitivity (0.44 vs 0.78; P = 0.25) and a similar negative predictive value (0.88 vs 0.86). All six DL-positive test nodules were molecular-positive, and the DL output split the molecular-positive nodules into a high-risk (4/6 malignant) and a lower-risk group (3/27 malignant; P = 0.01). Surgical verification was strongly driven by the molecular result (99\% of positive vs 19\% of negative results), and molecular-test accuracy depended on the platform. A logistic regression of the DL output, nodule size and the molecular result dropped the DL term and did not improve on the DL output alone. Ultrasound DL complemented molecular testing for Bethesda III nodules, in particular by identifying molecular-positive nodules at lower risk of malignancy.
\end{abstract}

\section{Introduction}

Thyroid nodules are a common clinical finding, and their evaluation often begins with ultrasound (US) imaging followed by fine needle aspiration (FNA) \citep{durante2018diagnosis}. This approach remains the gold standard for diagnosis and management \citep{alexander2022diagnosis}. However, despite its utility, the Bethesda System for Reporting Thyroid Cytopathology introduces a significant degree of uncertainty for indeterminate lesions, specifically those categorized as Bethesda III (Atypia of Undetermined Significance/Follicular Lesion of Undetermined Significance, AUS/FLUS) and Bethesda IV (Follicular Neoplasm/Suspicious for a Follicular Neoplasm) \citep{ho2014malignancy,kiernan2017bethesda}. These categories represent middle-risk lesions, with an estimated malignancy risk ranging from 5--15\% for Bethesda III and 15--30\% for Bethesda IV \citep{cibas2009bethesda}; later editions of the Bethesda System revised these estimates upwards on the basis of newer surgical follow-up data, whereas counting noninvasive follicular thyroid neoplasm with papillary-like nuclear features (NIFTP) as non-malignant lowers them again (for Bethesda III, 6--18\% rather than about 10--30\%) \citep{cibas2017bethesda,ali2023bethesda,nikiforov2016nomenclature}. For these nodules, clinicians face a critical decision: to proceed with surgery or to adopt a more conservative monitoring approach. This decision-making process is fraught with challenges due to the inherent ambiguity in cytological interpretation and the potential consequences of unnecessary surgical intervention \citep{iskandar2015evidence}.

In response to these diagnostic challenges, molecular testing has emerged as a transformative adjunct to cytopathology \citep{livhits2021effectiveness,roth2018molecular}, and current guidelines allow it to supplement the risk assessment of AUS/FLUS nodules \citep{haugen2016american}. Commercially available molecular assays, such as Thyroseq and Afirma, have been designed to further stratify the risk of malignancy in indeterminate thyroid nodules \citep{nikiforova2013targeted,mciver2014independent}. These tests leverage advanced genetic and transcriptomic analyses to classify nodules into low-risk categories, which can be monitored through serial US imaging, or high-risk categories, which may recommend surgeries \citep{nikiforova2018analytical,alexander2014multicenter}. Thyroseq utilizes next-generation sequencing to analyze genetic mutations, gene fusions, and other alterations associated with thyroid cancer \citep{nikiforov2014highly}. Similarly, Afirma employs a gene expression classifier to assess mRNA expression profiles, providing a binary classification of benign or suspicious nodules \citep{vuong2021diagnostic}. Clinical validation studies have demonstrated the robust performance of these molecular assays, reporting sensitivities and specificities that enhance the diagnostic confidence for indeterminate nodules \citep{steward2019performance,patel2018performance}, although single-institution experience, including that of Jiang and colleagues at Duke University, has reported more modest predictive values \citep{jug2018molecular}. However, despite their value, molecular tests have limitations, including high costs, accessibility barriers, and the potential for inconclusive results in a subset of cases \citep{keutgen2013molecular,mciver2014independent,khan2020thyroid,jassal2025beyond}.

Concurrently, artificial intelligence (AI) has gained traction as a powerful tool in medical imaging, particularly for the evaluation of thyroid nodules \citep{zhou2020differential}. Convolutional neural networks (CNNs), a class of AI algorithms designed for image analysis, have been extensively studied and validated in the context of thyroid ultrasound \citep{koh2020diagnosis}. CNNs have demonstrated exceptional performance in classifying thyroid nodules as benign or malignant prior to FNA, matching human radiologists and in some studies surpassing them in specificity \citep{buda2019management,li2019diagnosis,koh2020diagnosis}. For instance, recent studies have shown that deep learning models can effectively analyze ultrasound images to identify subtle morphological features associated with malignancy, such as irregular margins, microcalcifications, and taller-than-wide shape \citep{chen2022artificial,han2024automatic}. These features, when interpreted by CNNs, can be used to generate highly accurate risk assessments, providing a valuable decision aid for clinicians \citep{srivastava2023optimizing}. However, current AI tools primarily focus on guiding the decision to perform an FNA, and comparatively few address the diagnostic uncertainty that arises after an indeterminate FNA result \citep{zhu2019deep,jassal2025beyond}.

The studies that do address indeterminate nodules are small and heterogeneous. Daniels \etal trained an automated classifier on ultrasound images of 134 Bethesda III/IV nodules to predict the genetic risk category assigned by molecular testing \citep{daniels2020machine}. Gild \etal stratified 88 resected Bethesda III nodules and also evaluated a model trained mostly on determinate nodules \citep{gild2022risk}. Keutgen \etal reported a cross-validated area under the receiver operating characteristic curve (AUC) of 0.67 in indeterminate nodules and compared their algorithm with a commercial molecular test on a small independent set \citep{keutgen2022machine}. Kim \etal applied an existing deep learning model to 165 Bethesda III nodules from five institutions and reported an AUC of 0.75, with a sensitivity of 0.91 and a negative predictive value of 0.87 \citep{kim2026artificial}. CNN, radiomics and deep-learning nomogram models have also been proposed for AUS/FLUS and other indeterminate nodules \citep{youn2021diagnosing,yoon2021implications,zhong2025ultrasound}, including a multimodal model for Bethesda IV nodules \citep{ding2026multimodal}. Recently, Athreya \etal combined ultrasound deep learning with molecular test results in a multimodal model \citep{athreya2025combining}, and a systematic review concluded that the evidence for integrating AI-based ultrasound with molecular testing remains limited, with verification bias the most common risk of bias \citep{cortes2026integration}.

This study aims to address this gap by hypothesizing that CNN-based AI models can be developed to stratify indeterminate nodules (Bethesda III) into low- and high-risk categories, similar to molecular testing. By leveraging the strength of CNNs in analyzing ultrasound images, our approach seeks to offer a non-invasive, cost-effective, and widely accessible alternative to molecular assays. Such a tool could empower clinicians to make more informed decisions about whether to proceed with surgery or adopt a monitoring strategy for indeterminate nodules. The potential integration of CNNs into clinical workflows could complement existing diagnostic modalities and help optimize patient outcomes, particularly in settings where molecular testing may not be readily available or affordable. By bridging the gap between pre- and post-FNA diagnostic tools, this study represents a significant step forward in leveraging AI to enhance thyroid nodule management.

We compare the deep learning (DL) model with molecular testing on the same held-out nodules, and we test whether combining the two improves the prediction, both with regression models and with simple rules that need no fitting. We also examine how verification bias and the molecular testing platform affect the accuracy of molecular testing in our data, and how robust our comparison is to these and other factors. The DL model is intended for nodules with an AUS/FLUS result on FNA, either in place of molecular testing or together with it.

\section{Materials and Methods}
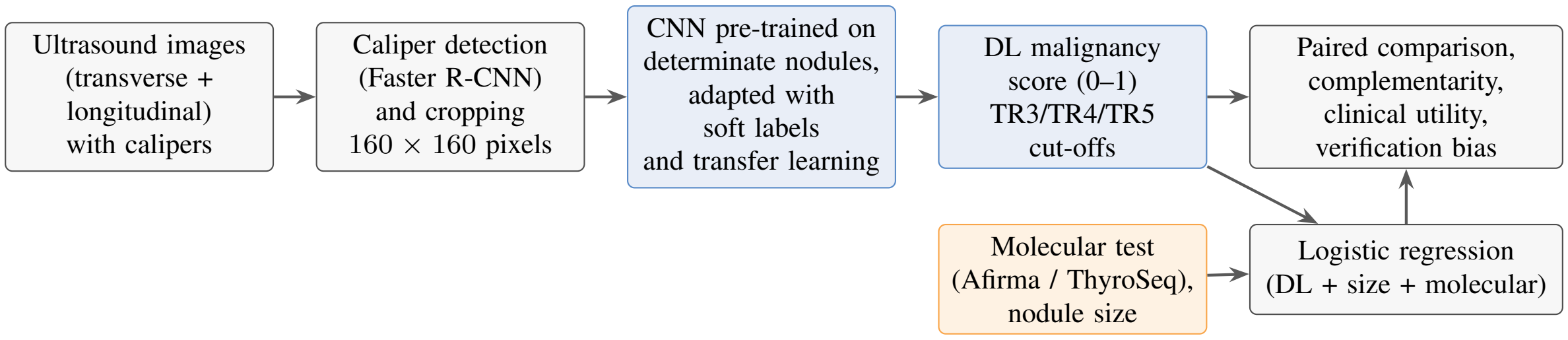
\begin{figure*}[t]
\centering
\begin{tikzpicture}[
  box/.style={draw=black!70, rounded corners=2pt, fill=black!3, align=center, font=\footnotesize, minimum height=1.25cm, text width=2.45cm},
  hl/.style={box, fill=NavyBlue!8, draw=NavyBlue!70},
  mol/.style={box, fill=BurntOrange!10, draw=BurntOrange!80},
  arr/.style={-{Stealth[length=2mm]}, thick, black!65},
  node distance=0.42cm]
\node[box] (us) {Ultrasound images\\(transverse + longitudinal)\\with calipers};
\node[box, right=of us] (det) {Caliper detection\\(Faster R-CNN)\\and cropping\\$160\times160$ pixels};
\node[hl, right=of det] (cnn) {CNN pre-trained on\\determinate nodules,\\adapted with soft labels\\and transfer learning};
\node[hl, right=of cnn] (score) {DL malignancy\\score (0--1)\\TR3/TR4/TR5\\cut-offs};
\node[box, right=of score, text width=2.9cm] (eval) {Paired comparison,\\complementarity,\\clinical utility,\\verification bias};
\node[mol, below=0.55cm of score, minimum height=0.9cm] (molbox) {Molecular test\\(Afirma / ThyroSeq),\\nodule size};
\node[box, below=0.55cm of eval, text width=2.9cm, minimum height=0.9cm] (lr) {Logistic regression\\(DL + size + molecular)};
\draw[arr] (us) -- (det);
\draw[arr] (det) -- (cnn);
\draw[arr] (cnn) -- (score);
\draw[arr] (score) -- (eval);
\draw[arr] (score) -- (lr);
\draw[arr] (molbox) -- (lr);
\draw[arr] (lr) -- (eval);
\end{tikzpicture}
\caption{Overview of the study pipeline. Each nodule's ultrasound images are cropped around the technologist's calipers and scored by a CNN adapted to indeterminate nodules. The DL score is compared with, and combined with, the molecular test result (and nodule size) of the same nodule against surgical pathology.}
\label{fig:pipeline}
\end{figure*}

This retrospective study was approved by our institutional review board and was HIPAA compliant. A waiver of informed consent was also obtained. The overall pipeline is summarized in Figure~\ref{fig:pipeline}.

\subsection{Study Population}
This study included patients with thyroid nodules classified as Atypia of Undetermined Significance/Follicular Lesion of Undetermined Significance (AUS/FLUS; Bethesda III) \citep{cibas2017bethesda} on cytological evaluation who underwent molecular testing between 2011 and 2022. Cases were identified through a retrospective review of institutional pathology and molecular testing databases. No other exclusion criteria were applied at this stage, ensuring a comprehensive dataset representing all eligible AUS/FLUS cases within the specified timeframe. The cases formed a consecutive series from a single academic medical center. In total, 466 nodule records from 418 patients were identified (Table~\ref{tab:demographics}), and 209 of them had a definitive surgical pathology diagnosis, 191 of which also had ultrasound images. Three patients (four nodules) were younger than 18 years; two of these nodules were in the analysis cohort, one of them in the test set.

Molecular testing was performed with Afirma (gene expression classifier, GEC, or genomic sequencing classifier, GSC) or ThyroSeq (v1, v2 or v3, or the earlier 7-gene mutation panel; the version was not recorded for one test nodule), and each result was dichotomized as positive (Afirma suspicious; ThyroSeq positive) or negative; ThyroSeq results reported as negative were counted as negative even when a low-risk alteration was detected. The reference standard was the diagnosis on surgical pathology. The diagnosis was taken from the clinical pathology report. We chose surgical pathology because it provides the definitive diagnosis for these nodules, and we did not assume that nodules without surgery were benign. Nodules diagnosed as NIFTP, and benign nodules with an incidental papillary microcarcinoma elsewhere in the gland, were counted as benign; counting NIFTP as malignant was examined in a sensitivity analysis. Two nodules with an inconclusive surgical pathology result were treated as unverified, and one recorded as possibly malignant was counted as malignant. The DL scores were generated retrospectively and were not available to the treating clinicians or the pathologists. The molecular result was part of the clinical record.

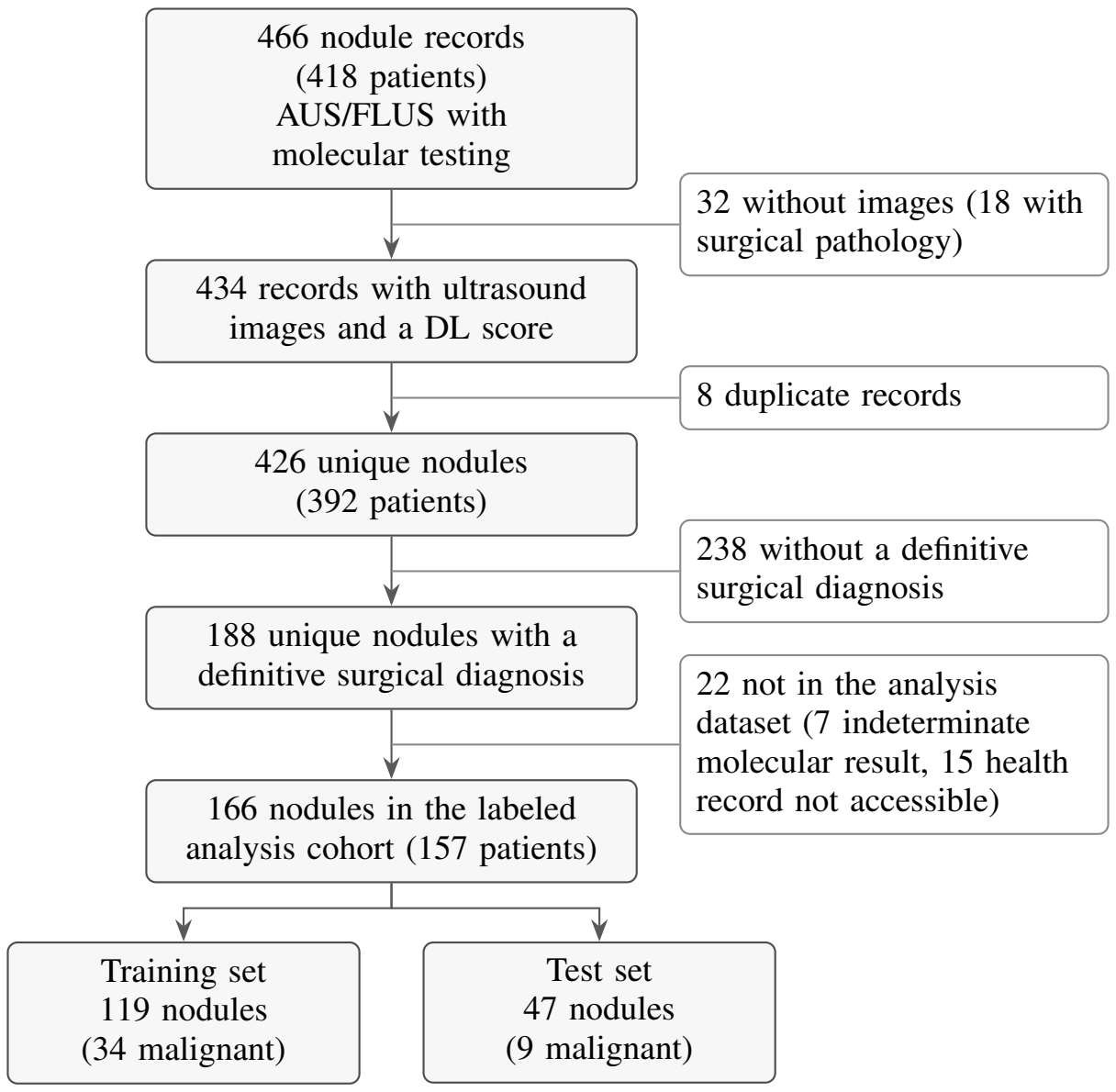
\begin{figure}[t]
\centering
\begin{tikzpicture}[
  fb/.style={draw=black!70, rounded corners=2pt, fill=black!3, align=center, font=\scriptsize, text width=3.3cm, minimum height=0.62cm},
  side/.style={draw=black!45, rounded corners=2pt, fill=white, align=left, font=\scriptsize, text width=2.85cm},
  arr/.style={-{Stealth[length=1.6mm]}, black!65},
  node distance=0.5cm]
\node[fb] (a) {466 nodule records (418 patients)\\AUS/FLUS with molecular testing};
\node[fb, below=of a] (b) {434 records with ultrasound\\images and a DL score};
\node[fb, below=of b] (c) {426 unique nodules\\(392 patients)};
\node[fb, below=of c] (d) {188 unique nodules with a\\definitive surgical diagnosis};
\node[fb, below=of d] (e) {166 nodules in the labeled\\analysis cohort (157 patients)};
\node[fb, below left=0.42cm and -1.55cm of e, text width=2.3cm] (tr) {Training set\\119 nodules\\(34 malignant)};
\node[fb, below right=0.42cm and -1.55cm of e, text width=2.3cm] (te) {Test set\\47 nodules\\(9 malignant)};
\coordinate (sidex) at ($(a.east)+(0.3,0)$);
\coordinate (m1) at ($(a.south)!0.5!(b.north)$);
\coordinate (m2) at ($(b.south)!0.5!(c.north)$);
\coordinate (m3) at ($(c.south)!0.5!(d.north)$);
\coordinate (m4) at ($(d.south)!0.5!(e.north)$);
\node[side, anchor=west] (x1) at (sidex |- m1) {32 without images (18 with surgical pathology)};
\node[side, anchor=west] (x2) at (sidex |- m2) {8 duplicate records};
\node[side, anchor=west] (x3) at (sidex |- m3) {238 without a definitive surgical diagnosis};
\node[side, anchor=west] (x4) at (sidex |- m4) {22 not in the analysis dataset (7 indeterminate molecular result, 15 health record not accessible)};
\draw[arr] (a) -- (b);
\draw[arr] (b) -- (c);
\draw[arr] (c) -- (d);
\draw[arr] (d) -- (e);
\draw[arr] (e.south) -- ++(0,-0.18) -| (tr.north);
\draw[arr] (e.south) -- ++(0,-0.18) -| (te.north);
\draw[black!45] (m1) -- (x1.west);
\draw[black!45] (m2) -- (x2.west);
\draw[black!45] (m3) -- (x3.west);
\draw[black!45] (m4) -- (x4.west);
\end{tikzpicture}
\caption{Flow of nodules through the study. Duplicate records (identical images entered under two identifiers) were removed before analysis; three of them were in the original testing dataset of 50 records.}
\label{fig:flow}
\end{figure}

Of the 466 nodule records, 434 had ultrasound images available and were scored by the DL algorithm. While preparing this manuscript, a review of the image files identified 8 nodules that had been entered twice under different identifiers (the same patient, with pixel-identical images and therefore identical DL outputs). The second record of each was removed, leaving 426 unique imaged nodules from 392 patients (Figure~\ref{fig:flow}). Of the 191 imaged nodule records with surgical pathology, 169 formed the labeled analysis dataset of this study; 7 nodules whose molecular result was indeterminate or unavailable were not part of this dataset, and 15 nodules were removed because their electronic health records could not be accessed. The testing dataset consisted of the last 50 records of the analysis dataset in order of record number, and the remaining records formed the training set. Three of the duplicate records were in the testing dataset: two were second records of training nodules and one was a second record of another test nodule. After their removal, the analysis cohort comprised 166 unique nodules from 157 patients, with 119 training nodules (34 malignant) and 47 test nodules (9 malignant), and no patient contributed nodules to both sets (Table~\ref{tab:split}). The 238 unique imaged nodules without a definitive surgical diagnosis (including the two with an inconclusive result) were used only in the correlation and verification-bias analyses.

\begin{table}[t]
\centering
\caption{Demographic Characteristics of the Study Population}
\begin{tabular}{l|l}
\hline
\textbf{Demographic} & \textbf{Total (N=466)} \\
\hline
\textbf{Sex} & \\
Female & 374 \\
Male & 82 \\
Unknown & 10 \\
\hline
\textbf{Age} & \\
Overall Mean $\pm$ SD (Range) & 57.23 $\pm$ 14.50 (6, 87) \\
Female Mean $\pm$ SD (Range) & 56.41 $\pm$ 14.94 (6, 87) \\
Male Mean $\pm$ SD (Range) & 60.96 $\pm$ 11.62 (32, 85) \\
\hline
\end{tabular}
\label{tab:demographics}
\end{table}

\begin{table}[t]
\centering
\caption{Characteristics of the training and test nodules (analysis cohort after removal of duplicate records). PTMC: incidental papillary thyroid microcarcinoma elsewhere in the gland. P values from Mann--Whitney U tests for age and size, chi-square tests for sex (female, male and unknown) and molecular platform, and Fisher's exact tests for the molecular result and for surgical pathology (malignant vs benign).}
\label{tab:split}
\footnotesize
\setlength{\tabcolsep}{2.6pt}
\resizebox{\linewidth}{!}{\begin{tabular}{lccc}
\toprule
\textbf{Characteristic} & \textbf{Training (n=119)} & \textbf{Test (n=47)} & \textbf{P} \\
\midrule
Age, years, mean $\pm$ SD & 55.3 $\pm$ 14.4 & 55.4 $\pm$ 15.7 & 0.92 \\
Female sex, n (\%) & 97 (81.5) & 39 (83.0) & 0.81 \\
Size, cm, median (IQR) & 2.0 (1.4--3.0) & 2.2 (1.6--2.8) & 0.34 \\
Molecular platform, n (\%) & & & $<$0.001 \\
\quad Afirma GEC & 81 (68.1) & 10 (21.3) & \\
\quad ThyroSeq 7-gene & 27 (22.7) & 0 (0.0) & \\
\quad ThyroSeq v2 & 11 (9.2) & 9 (19.1) & \\
\quad ThyroSeq v3 & 0 (0.0) & 27 (57.4) & \\
\quad ThyroSeq, version not recorded & 0 (0.0) & 1 (2.1) & \\
Molecular result, n (\%) & & & 0.59 \\
\quad Positive & 78 (65.5) & 33 (70.2) & \\
\quad Negative & 41 (34.5) & 14 (29.8) & \\
\quad Indeterminate & 0 (0.0) & 0 (0.0) & \\
Surgical pathology, n (\%) & & & 0.24 \\
\quad Malignant & 34 (28.6) & 9 (19.1) & \\
\quad Benign & 75 (63.0) & 31 (66.0) & \\
\quad Benign, incidental PTMC & 8 (6.7) & 4 (8.5) & \\
\quad NIFTP (benign) & 2 (1.7) & 3 (6.4) & \\
\bottomrule
\end{tabular}
}
\end{table}

\subsection{Data Preparation}
For each nodule, a single transverse and a single longitudinal grayscale static image were chosen. These images were carefully selected to provide the clearest depiction of the nodule, including its boundaries and caliper measurements placed by the technologist at the time of image acquisition. Technical parameters, such as pixel spacing, were preserved to allow for potential downstream analyses, but all patient-identifying information, including examination dates, age, and sex, was removed to ensure complete de-identification.

The selected images were initially stored in DICOM format, which was subsequently converted to PNG to standardize the data format and facilitate integration into the computational pipeline \citep{pianykh2012digital}. Uniform cropping was applied to all ultrasound images to remove all patient information.

A detection algorithm was employed to identify and localize the calipers marking the nodule boundaries to isolate individual thyroid nodules from the entire ultrasound images. The detection algorithm utilized a deep learning-based object detection framework, Faster R-CNN, trained on annotated ultrasound images with caliper markings in a previous study \citep{ren2017faster,buda2019management}. Once the calipers were detected, their coordinates were used to define a bounding box around the nodule. The algorithm automatically extracted this region of interest by cropping the bounding box from the original ultrasound image. This approach ensured that the cropped images contained the nodule and excluded irrelevant surrounding structures, such as adjacent tissues or background noise. The cropped images were then standardized in size to ensure compatibility with subsequent image analysis pipelines. Specifically, the detector used a ResNet-101 backbone pre-trained on natural images, and a square region extending 32 pixels beyond the caliper bounding box was cropped and resized to $160 \times 160$ pixels.

\subsection{Deep Learning Algorithm}
The development of the convolutional neural network (CNN) for indeterminate thyroid nodule classification began by establishing a baseline performance using a previously trained classification algorithm. This baseline performance provided a benchmark against which improvements could be assessed.

To develop an adapted algorithm for indeterminate thyroid nodules, we incorporated soft-label indeterminate thyroid nodules into the thyroid ultrasound dataset \citep{nguyen2014learning}. Soft labels were derived through a knowledge distillation approach, leveraging predictions from a more complex pre-trained model to guide the training of the new network \citep{hinton2015distilling}. This allowed the model to account for uncertainty in nodule classifications while preserving the diagnostic information embedded within indeterminate cases. The initial model weights were obtained from pre-trained networks trained on determinate thyroid nodules, which provided a strong starting point for transfer learning \citep{buda2019management}.

Subsequently, transfer learning techniques were employed to fine-tune the model on the expanded dataset containing both determinate and indeterminate thyroid nodules \citep{zhuang2020comprehensive}. This step ensured that the network adapted to the nuances of the indeterminate nodule dataset. To further enhance the model's generalizability and maintain performance consistency as new data became available, continual learning methods were incorporated. These methods allowed the model to integrate new knowledge incrementally without forgetting previously learned patterns.

The pre-trained network is the multi-task CNN of Buda \etal \citep{buda2019management}, trained on 1278 nodules from 1139 adult patients with conclusive cytologic or histologic diagnoses. It takes a $160 \times 160$ grayscale crop as input and consists of six convolutional layers ($3 \times 3$ kernels with 8, 12, 16, 24, 32 and 48 filters) and five max-pooling layers, followed by dropout and a sigmoid output that gives a malignancy score between 0 and 1; auxiliary outputs predict American College of Radiology Thyroid Imaging Reporting and Data System (ACR TI-RADS) descriptors \citep{tessler2017acr}. In the original work, the malignancy score was mapped to TI-RADS-equivalent risk levels by matching the distribution of expert TI-RADS levels; in the cross-validation analysis of the original code, this mapping gave cut-offs of 0.342 (TR3), 0.433 (TR4) and 0.617 (TR5) (the publication reports 0.210, 0.253 and 0.347, derived from the final model's training-set predictions). We used these cut-offs, which were defined independently of the present data, as operating points. The TR5 cut-off (0.617, written 0.62 below) was the primary operating point; a threshold of 0.5 was used for the exploratory ``prediction label'' analyses.

\subsection{Comparative Performance and Relationship Analysis of Deep Learning Predictions and Molecular Testing}
\label{compare-df-gen}
We collected both the deep neural network's predictions and the molecular testing results, and performed an intermediate evaluation of each as a single predictor in a univariable logistic regression on the labeled training set (119 nodules), using 5-fold cross-validation to compute the AUC \citep{rodriguez2009sensitivity,ling2003auc}.

To examine the relationship between deep learning predictions and molecular testing outcomes, we calculated the Point Biserial Correlation coefficient between the neural network's predictions (ranging continuously from 0 to 1) and molecular testing results (binary: 0 or 1) \citep{tate1954correlation}. Additionally, we computed the Pearson correlation coefficient after binarizing the neural network predictions using a threshold of 0.5 \citep{cohen2009pearson}. Both analyses were conducted on the entire training set, including both labeled and unlabeled data (357 nodules with a definitive molecular result). The results showed a Point Biserial Correlation coefficient of 0.12 and a Pearson correlation coefficient of 0.02, indicating little to no association between the two variables. These findings suggest that deep learning predictions and molecular testing outcomes are largely independent. Given their comparable predictive accuracy shown in Section~\ref{sec:dev-results} when used individually, we hypothesize that combining these variables could provide complementary information, potentially enhancing the overall model performance.

\begin{figure}[t]
\centering
\includegraphics[width=\linewidth]{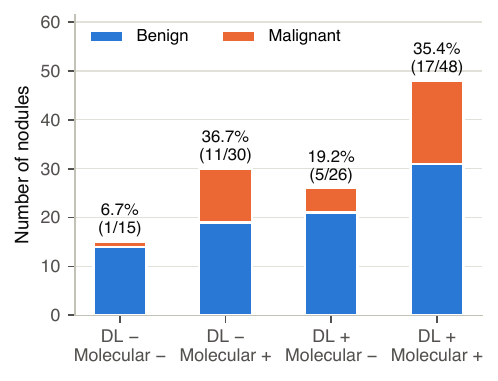}
\caption[Subgroup analysis showing the distribution of different combinations of prediction labels and molecular testing results on the training set.]{Subgroup analysis showing the distribution of different combinations of prediction labels and molecular testing results on the training set (n = 119; prediction label at a threshold of 0.5). The percentage displayed on top of each bar represents the proportion of gt=1 within each combination, calculated as gt=1 / (gt=1 + gt=0), with counts in parentheses.}
\label{fig:subgroups-train}
\end{figure}

Additionally, we compared the agreement between the deep learning (DL) prediction model, molecular testing, and the ground truth on the 119 labeled training nodules. To facilitate this analysis, we used a pseudo-threshold for grouping. Specifically, we categorized different combinations of prediction labels and molecular labels and calculated the percentage of positive cases (gt=1) within each group. As shown in Figure~\ref{fig:subgroups-train}, we observed that in 93.3\% of cases where both the prediction label and molecular testing indicated low risk (prediction label=0 and molecular=0), the ground truth was also negative (gt=0). This suggests that when both variables agree on a low-risk classification, the nodule is very likely in the low-risk group. However, in cases where both variables indicated high risk (the fourth group), only 35.4\% of these were actually positive (gt=1). This indicates that both the deep learning model and molecular testing tend to have heightened sensitivity, leading to an increased rate of false positives. Given that the prediction label represents a continuous probability and we tentatively applied a threshold of 0.5, we anticipate that fine-tuning this threshold could improve the accuracy of the high-risk group classification (the fourth group). Based on these findings, we hypothesize the following:
\begin{enumerate}
\item Combining the continuous prediction probabilities from the DL model with molecular testing results might enhance predictive accuracy.
\item Certain combinations of prediction labels and molecular testing results offer higher certainty about the ground truth. These combinations could be leveraged to generate soft labels for unlabeled data, potentially expanding and enriching the training set.
\end{enumerate}

\subsection{Development of Regression Model to Combine Deep Learning and Molecular Testing}
Previous analyses suggest that the deep learning (DL) model and molecular testing demonstrate comparable performance, with molecular testing showing a slight edge in the training set. Importantly, the two features were only weakly correlated, so each could contribute independently to predictive performance. Based on this, we decided to design a multivariable model that integrates both features \citep{nick2007logistic}. Additionally, we collected nodule size, considering it potentially helpful but also recognizing that it might provide redundant information since ultrasound images inherently capture this data.

To evaluate the contribution of these features, we developed several model variations with different combinations: (1) a DL prediction-only model; (2) DL predictions + nodule size; (3) DL predictions + molecular testing; and (4) DL predictions + nodule size + molecular testing. Logistic regression models were trained using 5-fold cross-validation on the training set to identify the optimal feature combination. As shown in the validation results (Table~\ref{tab:auc_scores}), the best-performing of these four models incorporated all three features, yielding an improvement in average validation AUC of approximately 0.08. For reference, we also evaluated models without the DL prediction (molecular testing, nodule size, and both), and repeated the cross-validation 200 times with different random partitions to quantify the variability of a single 5-fold split; models were compared over these partitions with the corrected resampled t-test \citep{nadeau2003inference}.

\begin{table*}[t]
\centering
\caption{Cross-validated AUC scores for different feature combinations (training set, n = 119; 5-fold cross-validation with a single seeded partition, and mean with 95\% range over 200 repeated partitions).}
\small
\begin{tabular}{|l|c|l|c|}
\hline
 \textbf{Features} & \textbf{Mean AUC $\pm$ Std} & \textbf{AUC Scores (Each Fold)} & \textbf{Repeated CV} \\ \hline
Prediction, nodsize, molecular & 0.6398 $\pm$ 0.0526 & 0.7395, 0.5966, 0.5966, 0.6387, 0.6275 & 0.663 (0.634--0.690) \\ \hline
Prediction, molecular & 0.6317 $\pm$ 0.0923 & 0.5882, 0.6134, 0.6387, 0.7983, 0.5196 & 0.628 (0.568--0.656) \\ \hline
Prediction, nodsize & 0.6272 $\pm$ 0.0997 & 0.8067, 0.5714, 0.5126, 0.6471, 0.5980 & 0.639 (0.617--0.662) \\ \hline
Prediction & 0.5583 $\pm$ 0.1104 & 0.5042, 0.5462, 0.5630, 0.7563, 0.4216 & 0.553 (0.524--0.576) \\ \hline
\multicolumn{4}{|l|}{\textit{Reference models without the DL prediction}} \\ \hline
Nodsize, molecular & 0.6471 $\pm$ 0.0545 & 0.7311, 0.5882, 0.5882, 0.6513, 0.6765 & 0.668 (0.639--0.694) \\ \hline
Molecular & 0.6178 $\pm$ 0.0276 & 0.5924, 0.6345, 0.6050, 0.6639, 0.5931 & 0.618 (0.613--0.620) \\ \hline
Nodsize & 0.6359 $\pm$ 0.0970 & 0.7941, 0.5672, 0.5084, 0.6429, 0.6667 & 0.645 (0.626--0.666) \\ \hline
\end{tabular}

\label{tab:auc_scores}
\end{table*}

After determining the optimal feature combination, we performed hyper-parameter tuning using these features on the entire training set. A grid search was conducted over the following parameter space: penalty (l1 or l2), classifier regularization strength (classifier\_c, ranging from np.logspace(-3, 3, 7)), and class weight (None or balanced) to account for potential class imbalance. The best model was ultimately selected with classifier\_c set to 1, l1 penalty, and balanced class weights. This model was then frozen and applied once to the test set.

\subsection{Statistical Analysis}
\label{sec:stats}
Diagnostic accuracy is reported following STARD 2015 \citep{bossuyt2015stard}. The held-out test set was used for the primary evaluation, the paired comparison of the DL model at the TR5 cut-off with molecular testing; the training set was used only for the exploratory and model-development analyses above. The sample size was set by the available data, and we did not perform a sample size calculation in advance. Sensitivity, specificity and positive and negative predictive values (PPV, NPV) are reported with exact Clopper--Pearson 95\% confidence intervals (CIs) \citep{clopper1934use}, and likelihood ratios (LR+, LR$-$) with log-method CIs for the primary comparison. The area under the receiver operating characteristic curve (AUC) is reported with DeLong CIs and, as a check, stratified bootstrap CIs (10{,}000 resamples) \citep{delong1988comparing,efron1994introduction}; for the binary molecular result, the AUC equals the mean of sensitivity and specificity \citep{hanley1982meaning}. Paired comparisons of the DL model and molecular testing on the same nodules used the DeLong test for AUCs, the exact McNemar test for sensitivity and specificity \citep{mcnemar1947note}, with Newcombe's score interval for the paired difference \citep{newcombe1998improved}, and a nodule-level bootstrap for differences in PPV and NPV (resamples without a DL-positive nodule, 11 of 10{,}000, were discarded). Malignancy rates in subgroups defined by the two tests were compared with Fisher's exact test.

Association between the two tests was summarized by the point-biserial coefficient, the phi coefficient (the Pearson correlation of the binarized outputs, Section~\ref{compare-df-gen}) and Cohen's $\kappa$. Because two tests that both track malignancy will be correlated even if their errors are independent, we also computed the correlation within benign and within malignant nodules (conditional dependence) \citep{vacek1985effect}. Clinical utility was assessed with decision-curve analysis of referral strategies \citep{vickers2006decision} and with predictive values recomputed across a range of malignancy prevalence using Bayes' theorem \citep{fagan1975nomogram}. Calibration of the DL output was summarized by the calibration intercept and slope \citep{vancalster2019calibration}.

Because molecular-negative nodules were much less likely to undergo surgery, we estimated verification-bias-corrected sensitivity and specificity of molecular testing with the Begg--Greenes method, which assumes that surgery depended only on the molecular result \citep{begg1983assessment,degroot2011verification}, using the 404 unique imaged nodules with a definitive molecular result that were in the analysis cohort or had no definitive surgical diagnosis, and applied the corresponding inverse-probability-of-verification weights to the test set \citep{alonzo2005assessing}. Further sensitivity analyses counted NIFTP as malignant \citep{nikiforov2016nomenclature}, stratified molecular-test accuracy by platform, and repeated the main comparison on the test set as originally assembled (50 records, including duplicates) and on all 188 unique nodules with surgical pathology. All tests were two-sided with a significance level of 0.05 and were not adjusted for multiple comparisons; analyses other than the primary comparison should be read as exploratory. Analyses used Python 3.12 with scikit-learn 1.7.1 \citep{pedregosa2011scikit}, SciPy 1.16.0 \citep{virtanen2020scipy} and statsmodels 0.15.0 \citep{seabold2010statsmodels}. A large language model was used for agentic tasks, such as writing and running the statistical analysis code, and for refining the manuscript text. The authors reviewed the results and the text and take full responsibility for the content.

\section{Results}
\subsection{Study Cohort}
The flow of nodules is shown in Figure~\ref{fig:flow}. The analysis cohort comprised 166 nodules (43 malignant, 25.9\%). The training and test sets did not differ significantly in age, sex, nodule size, molecular result or malignancy rate (28.6\% vs 19.1\%), but differed in the molecular platform (P $<$ 0.001; Table~\ref{tab:split}): Afirma GEC was used for 68.1\% of the training nodules, whereas ThyroSeq v3 was used for 57.4\% of the test nodules and for none of the training nodules. Across the analysis cohort, molecular testing had a sensitivity of 0.81 (35/43; 95\% CI 0.67--0.92), a specificity of 0.38 (47/123; 0.30--0.47), a PPV of 0.32 (35/111) and an NPV of 0.85 (47/55).

\subsection{Development-Set Analyses}
\label{sec:dev-results}
In the training set, the DL output had a univariable AUC of 0.56 (95\% CI 0.44--0.67) and molecular testing an AUC of 0.62 (0.53--0.70). At the TR5 cut-off, the DL output did not discriminate in the training set (sensitivity 0.29, specificity 0.71, LR+ 1.00); 10 of the 35 DL-positive training nodules were molecular-negative, and among molecular-positive training nodules the malignancy rate was the same with a positive or a negative DL result (9 of 25 vs 19 of 53). The correlation between the two was low (point-biserial 0.13 among the 119 labeled training nodules), and nodules negative on both tests at the 0.5 threshold were rarely malignant (1 of 15; Figure~\ref{fig:subgroups-train}). Nodule size was inversely associated with malignancy: smaller nodules were more often malignant (AUC 0.35 with size used directly as the score; the size-only regression in Table~\ref{tab:auc_scores} reverses the sign and reaches 0.64).

\subsection{Deep Learning Versus Molecular Testing Performance}
\begin{figure*}[t]
\centering
\includegraphics[width=0.82\linewidth]{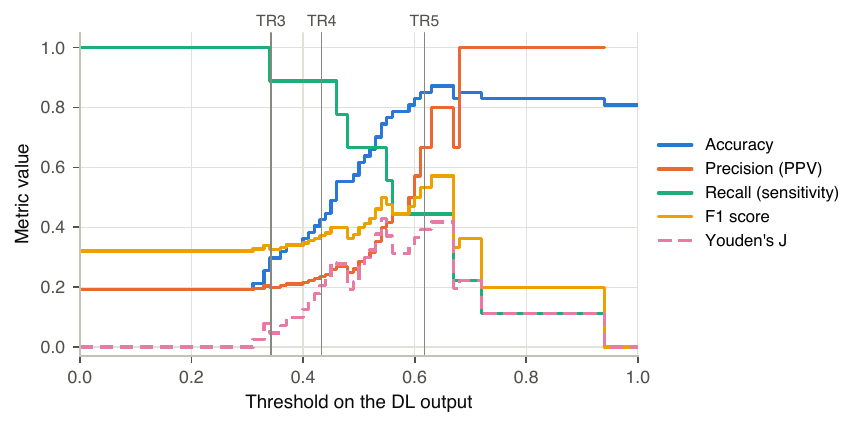}
\caption{Different metric results for deep learning prediction only on the test set (n = 47). Vertical lines mark the TR3, TR4 and TR5 cut-offs inherited from the original algorithm.}
\label{fig:dl-only}
\end{figure*}

In our earlier analysis of this cohort \citep{yang2024machine}, the original deep learning model was reported with a sensitivity of 0.35, a specificity of 0.78 and an NPV of 0.77 at a threshold of 0.62 on all 191 labeled nodule records, including the test nodules; on the same records, molecular testing had a sensitivity of 0.78, a specificity of 0.42 and an NPV of 0.84.

On the testing set, our fine-tuned deep learning model achieved a sensitivity of 0.44 (Figure~\ref{fig:dl-only}) and a specificity of 0.95, with a negative predictive value (NPV) of 0.88. Molecular testing had a sensitivity of 0.78 and a specificity of 0.32, with an NPV of 0.86.

\begin{figure}[t]
\centering
\includegraphics[width=0.92\linewidth]{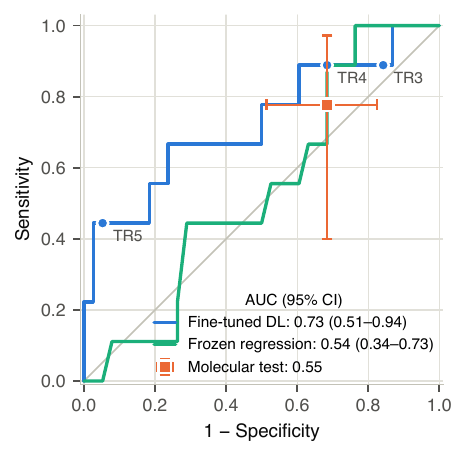}
\caption{Receiver operating characteristic curves on the test set (n = 47). Dots mark the TR3--TR5 cut-offs of the DL model; the square marks the molecular test with exact 95\% CIs for sensitivity and 1 $-$ specificity. The frozen regression model was fitted on the training set to the DL output, nodule size and the molecular result; the l1 penalty set the DL weight to zero.}
\label{fig:roc}
\end{figure}

Confidence intervals, predictive values and likelihood ratios are listed in Table~\ref{tab:test}. The DL model was more specific than molecular testing (36 vs 12 of the 38 benign nodules correctly classified; McNemar P $<$ 0.001) and had a higher PPV (0.67 vs 0.21; difference 0.45, 95\% CI 0.07--0.84), whereas the differences in sensitivity (4 vs 7 of 9 malignant nodules; P = 0.25) and NPV (0.88 vs 0.86) were not significant (Table~\ref{tab:paired}). A positive DL result carried an LR+ of 8.44 (95\% CI 1.82--39.15), compared with 1.14 (0.75--1.71) for a positive molecular result; the LR$-$ was 0.59 (0.33--1.06) and 0.70 (0.19--2.60), respectively. Across all thresholds, the DL model achieved an AUC of 0.73 (95\% CI 0.51--0.94; bootstrap 0.51--0.91), compared with 0.55 (0.38--0.71) for molecular testing (DeLong P = 0.15; Figure~\ref{fig:roc}). At the more permissive TR4 cut-off, the DL model detected 8 of the 9 malignant nodules (sensitivity 0.89) at the same specificity as molecular testing (0.32).

\begin{table*}[t]
\centering
\caption{Diagnostic performance on the held-out test set (n = 47, 9 malignant), with exact 95\% CIs. TP/FP/FN/TN: numbers of true-positive, false-positive, false-negative and true-negative results. $^{\ast}$Primary operating point.}
\label{tab:test}
\resizebox{\textwidth}{!}{\begin{tabular}{lcccccc}
\toprule
\textbf{Index test} & \textbf{TP/FP/FN/TN} & \textbf{Sensitivity} & \textbf{Specificity} & \textbf{PPV} & \textbf{NPV} & \textbf{LR+ / LR$-$} \\
\midrule
DL $>$ TR3 cut-off (0.342) & 8/32/1/6 & 0.89 (0.52--1.00) & 0.16 (0.06--0.31) & 0.20 (0.09--0.36) & 0.86 (0.42--1.00) & 1.06 / 0.70 \\
DL $>$ TR4 cut-off (0.433) & 8/26/1/12 & 0.89 (0.52--1.00) & 0.32 (0.18--0.49) & 0.24 (0.11--0.41) & 0.92 (0.64--1.00) & 1.30 / 0.35 \\
DL $>$ TR5 cut-off (0.617)$^{\ast}$ & 4/2/5/36 & 0.44 (0.14--0.79) & 0.95 (0.82--0.99) & 0.67 (0.22--0.96) & 0.88 (0.74--0.96) & 8.44 / 0.59 \\
Molecular test & 7/26/2/12 & 0.78 (0.40--0.97) & 0.32 (0.18--0.49) & 0.21 (0.09--0.39) & 0.86 (0.57--0.98) & 1.14 / 0.70 \\
Frozen regression model ($p \geq 0.5$) & 5/21/4/17 & 0.56 (0.21--0.86) & 0.45 (0.29--0.62) & 0.19 (0.07--0.39) & 0.81 (0.58--0.95) & 1.01 / 0.99 \\
\bottomrule
\end{tabular}
}
\end{table*}

\begin{table*}[t]
\centering
\caption{Paired comparison of the DL model (TR5 cut-off) and molecular testing on the 47 test nodules. Discordant pairs are given as DL-only correct : molecular-only correct.}
\label{tab:paired}
\small
\begin{tabular}{lcccl}
\toprule
\textbf{Measure} & \textbf{DL (TR5)} & \textbf{Molecular} & \textbf{Difference (95\% CI)} & \textbf{P (test)} \\
\midrule
AUC & 0.728 & 0.547 & 0.181 ($-$0.063 to 0.426) & 0.15 (DeLong) \\
Sensitivity & 4/9 & 7/9 & $-$0.33 ($-$0.60 to 0.05) & 0.25 (McNemar; discordant 0:3) \\
Specificity & 36/38 & 12/38 & 0.63 (0.44 to 0.76) & $<$0.001 (McNemar; discordant 24:0) \\
PPV & 4/6 & 7/33 & 0.45 (0.07 to 0.84) & 0.03 (bootstrap) \\
NPV & 36/41 & 12/14 & 0.02 ($-$0.12 to 0.19) & 0.82 (bootstrap) \\
\bottomrule
\end{tabular}

\end{table*}

\subsection{Correlation of Deep Learning Results and Molecular Testing Results}
We also examined the correlation between deep learning predictions and molecular testing results in the test set. The analysis revealed a point-biserial correlation coefficient of 0.33 (P = 0.02) for continuous predictions versus binary molecular testing outcomes, and a Pearson correlation coefficient of 0.30 (P = 0.04) for binary deep learning predictions and molecular testing. These findings indicate that, on the test set, the two prediction methods exhibit a weak-to-moderate positive correlation, stronger than in the training data. Within the benign test nodules the correlation persisted (point-biserial 0.37, P = 0.02), and it was of similar size within the nine malignant nodules (0.28, P = 0.47). Cohen's $\kappa$ at the TR5 cut-off was 0.12, the highest value possible given that 6 and 33 nodules were positive on the two tests.

\begin{figure*}[t]
\centering
\includegraphics[width=0.9\linewidth]{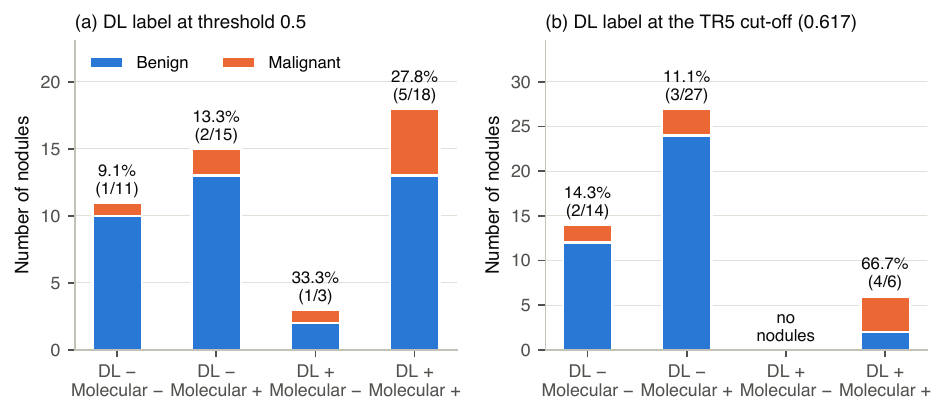}
\caption[Subgroup analysis showing the distribution of different combinations of prediction labels and molecular testing results on the test set.]{Subgroup analysis showing the distribution of different combinations of prediction labels and molecular testing results on the test set (n = 47), with the prediction label defined at (a) a threshold of 0.5 and (b) the TR5 cut-off. The percentage displayed on top of each bar represents the proportion of gt=1 within each combination, calculated as gt=1 / (gt=1 + gt=0).}
\label{fig:subgroups-test}
\end{figure*}

The subgroup analysis of various combinations of deep learning and molecular testing results on the test set reveals a trend consistent with the training set: (1) both methods exhibit stronger agreement in identifying low or no-risk groups, and (2) in Groups 2 and 3, disagreements between the models make it less evident which prediction is more reliable (Figure~\ref{fig:subgroups-test}a).

At the TR5 cut-off (Figure~\ref{fig:subgroups-test}b), all six DL-positive test nodules were also molecular-positive. Among the 33 molecular-positive nodules, 4 of the 6 DL-positive nodules (66.7\%) were malignant, compared with 3 of the 27 DL-negative nodules (11.1\%; odds ratio 16.0, Fisher P = 0.01). The malignancy rate of molecular-positive, DL-negative nodules was thus similar to that of molecular-negative nodules (2 of 14, 14.3\%). The three malignant nodules among the molecular-positive, DL-negative group were a follicular variant of papillary carcinoma with \textit{NRAS} and \textit{TERT} promoter co-mutations, a minimally invasive follicular carcinoma with an \textit{EIF1AX} mutation, and an oncocytic (H\"urthle cell) carcinoma with an \textit{EIF1AX} mutation; the two molecular-negative malignancies were multifocal papillary carcinomas.

\subsection{Combining Deep Learning and Molecular Testing}
As shown in Table~\ref{tab:auc_scores}, the best-performing regression model among those that included the DL prediction used the combination of deep learning prediction, nodule size and molecular testing result, with a mean 5-fold validation AUC of 0.64. Over 200 repeated partitions, its mean AUC was 0.66 (95\% range of repeat means 0.63--0.69). A reference model without the DL prediction (nodule size and molecular testing) performed similarly (0.65 on the single partition and 0.67 over repeated partitions), indicating that in the training set most of the gain came from molecular testing and nodule size. After hyper-parameter tuning, the l1 penalty set the coefficient of the DL prediction to zero, so that the frozen model reduced to nodule size and molecular testing. Applied once to the test set, it achieved an AUC of 0.54 (95\% CI 0.34--0.73), lower than that of the DL output alone (difference $-$0.19, DeLong P = 0.12; Figure~\ref{fig:roc}).

Combining the two binary results with simple rules that require no fitting illustrates the trade-off between surgeries and missed cancers (Supplementary Table~\ref{tab:strategies}). Operating only on nodules that were both molecular-positive and above the TR4 cut-off would have referred 26 instead of 33 nodules to surgery (20 instead of 26 benign), at the cost of one additional missed malignancy (3 vs 2). Conversely, nodules negative on both tests at the TR4 cut-off (6 of 47) were all benign.

\subsection{Clinical Utility}
\begin{figure*}[t]
\centering
\includegraphics[width=\linewidth]{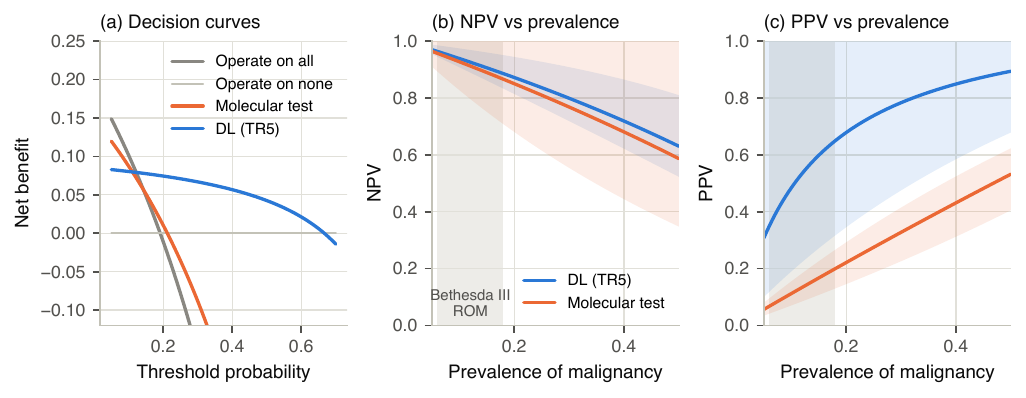}
\caption{Clinical utility on the test set. (a) Decision curves for referral to surgery based on the DL model (TR5 cut-off), the molecular test, all nodules or none. (b, c) NPV and PPV implied by the sensitivity and specificity of the resected test nodules (not weighted for verification) as a function of the prevalence of malignancy, with bootstrap 95\% bands; the shaded range is the risk of malignancy (ROM) of AUS/FLUS when NIFTP is not counted as cancer (6--18\%) \citep{cibas2017bethesda}.}
\label{fig:utility}
\end{figure*}

In decision-curve analysis (Figure~\ref{fig:utility}a), referral based on the DL model at the TR5 cut-off had the highest net benefit for threshold probabilities between approximately 12\% and 67\%, that is, when a clinician would accept between one half and about seven operations on benign nodules for each cancer detected. At a threshold probability of 20\%, the net benefit was 0.074 for the DL model, 0.011 for molecular testing and $-$0.011 for operating on all nodules, corresponding to a net reduction of 34 and 8.5 unnecessary operations per 100 nodules, respectively, compared with operating on all; in absolute terms, referral based on the DL model would have spared 36 of the 38 benign test nodules an operation but missed 5 of the 9 cancers. Because surgically verified cohorts overestimate the malignancy risk of AUS/FLUS nodules \citep{cibas2017bethesda}, we also recomputed predictive values across prevalences (Figure~\ref{fig:utility}b,c). At a prevalence of 15\%, the implied NPV was 0.91 for the DL model and 0.89 for molecular testing, and the implied PPV was 0.60 and 0.17, respectively.

\subsection{Verification Bias and Sensitivity Analyses}
Among the 404 nodules of the verification-bias analysis, surgical verification depended strongly on the molecular result: 111 of 112 molecular-positive nodules (99.1\%), but only 55 of 292 molecular-negative nodules (18.8\%), had a definitive surgical diagnosis. After Begg--Greenes correction, the estimated sensitivity of molecular testing fell from 0.81 to 0.45 (95\% CI 0.30--0.69) and its specificity rose from 0.38 to 0.76 (0.71--0.81), with an estimated malignancy prevalence of 19.3\% among these nodules (Supplementary Table~\ref{tab:robust}). Weighting the test set accordingly lowered the sensitivity of both tests but preserved the specificity advantage of the DL model (0.98 vs 0.71). On resected nodules, molecular-test accuracy differed markedly by platform: Afirma GEC was positive in all 25 malignant nodules but also in 53 of 66 benign nodules (sensitivity 1.00, specificity 0.20), whereas ThyroSeq had a sensitivity of 0.56 and a specificity of 0.60 (Fisher P $<$ 0.001 for both). Counting NIFTP as malignant changed the DL AUC from 0.73 to 0.70.

On the test set as originally assembled (50 records), results were similar to the primary analysis (DL AUC 0.76; specificity 0.95 vs 0.30). When all 188 unique nodules with surgical pathology were analyzed together, irrespective of the training/test assignment and including the 22 nodules outside the analysis dataset, the DL model and molecular testing had similar AUCs (0.59 and 0.60, P = 0.78); the DL model remained more specific (0.78 vs 0.43, P $<$ 0.001) but was less sensitive (0.34 vs 0.78, P $<$ 0.001), and the separation of molecular-positive nodules was weaker (13 of 31 vs 26 of 87 malignant, P = 0.27) (Supplementary Table~\ref{tab:sets}). The raw DL output was not calibrated to this population (mean 0.49 on the test set against an observed malignancy rate of 0.19; calibration intercept $-$1.52, 95\% CI $-$2.27 to $-$0.76; slope 2.04, 0.13--3.94), so it should be read as a ranking score rather than a calibrated probability.

\section{Discussion}
The ability to accurately classify indeterminate thyroid nodules represents a significant challenge in clinical practice. Current approaches, such as molecular testing, have demonstrated substantial utility in risk stratification. However, molecular testing is not without limitations, including high costs, variable accessibility, and imperfect diagnostic performance \citep{khan2020thyroid,fung2025high,jassal2025beyond}. In our analysis cohort, molecular testing achieved a sensitivity of 0.81, a specificity of 0.38, and an NPV of 0.85. These barriers underscore the need for additional tools to guide the management of indeterminate nodules, particularly tools that are widely accessible and cost-effective.

The original algorithm was trained on determinate thyroid nodules, using nodules with conclusive benign or malignant cytologic or histologic diagnoses, so indeterminate nodules, namely cytologically AUS/FLUS (Bethesda III) in our dataset, were outside its training data; this motivated the fine-tuning. On the test set, the fine-tuned model surpassed the molecular testing in specificity and PPV. To properly manage indeterminate thyroid nodules, negative predictive value (NPV) is a key index. To eliminate unnecessary surgeries, it is important to be certain that what we label as low risk should actually be low risk. The fine-tuned model has an NPV of 0.88, which is about the same level as the molecular testing (0.86); neither value reaches the NPV of about 96\% expected of a rule-out test, that of a benign cytology result \citep{haugen2016american}. All things considered, it is safe to conclude that in our dataset the fine-tuned model performs comparably to the molecular testing, with markedly higher specificity but lower sensitivity. The new algorithm has several advantages compared to molecular testing: it is more accessible to patients, it has very low cost, and its results are available instantly.

The two tests work in different ways. Molecular testing was sensitive but not specific: 26 of the 38 benign test nodules were molecular-positive, and a positive result barely changed the odds of malignancy (LR+ 1.14). The DL model at the TR5 cut-off was highly specific, and a DL-positive result increased the odds of malignancy about eightfold. Because every DL-positive test nodule was also molecular-positive, the DL model split the molecular-positive nodules into a high-risk group (4 of 6 malignant) and a lower-risk group (3 of 27 malignant, 11\%), close to the risk of molecular-negative nodules. This suggests using the DL score after a positive molecular result, for example to support surveillance for molecular-positive, DL-negative nodules. However, this strategy would have missed three of the seven molecular-positive cancers in our test set, one of which carried an \textit{NRAS} mutation together with a \textit{TERT} promoter mutation, the latter associated with aggressive disease \citep{melo2014tert}. DL-guided de-escalation should therefore be limited to molecular results without high-risk alterations; after excluding results with alterations rated high risk, 1 of the remaining 22 molecular-positive, DL-negative test nodules was malignant (an \textit{EIF1AX}-mutated carcinoma rated intermediate-high risk).

Comparing the fine-tuned model with molecular testing, we found that their correlation was low in the development data (0.12--0.13) and weak to moderate on the test set (0.33). Given that both have reasonable performance, we speculate that they rely on partially different information. Hence, theoretically, we can combine the two practices to refine the prediction and potentially increase the performance. In the 5-fold validation we performed, combining model prediction, nodule size, and molecular testing achieved an average AUC of 0.64, which is insignificantly higher than using model prediction alone or using molecular testing alone (corrected resampled t-test P = 0.11 and 0.36). On the test set, the frozen regression model did not improve on the DL output alone. Its weights were learned from training-set DL outputs that were less informative (AUC 0.56), so the l1 penalty removed the DL term, whereas on the test set the DL output was more informative (AUC 0.73). A combined model should therefore be trained on a larger labeled set, as in the multimodal model of Athreya \etal \citep{athreya2025combining}.

The accuracy of molecular testing in our cohort was lower than in the validation studies. In multicenter validation, ThyroSeq v3 reached a sensitivity of 91\% and a specificity of 85\% in Bethesda III nodules \citep{steward2019performance}, and the Afirma GSC 93\% and 71\% \citep{patel2018performance}; the earlier Afirma GEC had a specificity of 52\% across Bethesda III--V nodules \citep{alexander2012preoperative}. Our estimates on resected nodules are in line with post-marketing experience. For Afirma GEC (sensitivity 1.00, specificity 0.20), specificity on resected nodules was 25\% in a meta-analysis of 19 studies \citep{valderrabano2019comparison} and 0.26 in a later pooled analysis \citep{digennaro2022assessing}. For ThyroSeq v3 (0.83 and 0.33 in our cohort), pooled estimates were 0.92 and 0.41 \citep{digennaro2022assessing}; ThyroSeq v2 (0.50 and 0.56) was reported to be uninformative in Bethesda III nodules at one center \citep{valderrabano2017thyroseq}, and the earlier 7-gene mutation panel (0.38 and 0.89) behaved as a rule-in test \citep{haugen2016american}. Performance also varies substantially between institutions \citep{marti2015wide,marcadis2019interinstitutional}. A major reason for the low specificity on resected nodules is verification bias. In our cohort, almost all molecular-positive but fewer than one in five molecular-negative nodules were resected, a pattern also reported for Afirma GEC (80\% vs 20\%) \citep{valderrabano2019comparison}. Under this selection, the apparent sensitivity of molecular testing is inflated and its specificity deflated, while its predictive values are not affected as long as surgery depends only on the molecular result \citep{begg1983assessment,kohn2013understanding,degroot2011verification}. Once corrected, the specificity of molecular testing (0.76) approached reported validation values, and its sensitivity (0.45) fell well below them. Validation studies of these assays have generally not corrected for partial verification \citep{digennaro2022assessing,duh2017systematic}, and the same issue affects AI studies \citep{cortes2026integration}. Because the DL score was never available to the treating clinicians, verification did not depend on it (41\% of DL-positive and 41\% of DL-negative nodules had a definitive surgical diagnosis). It did depend on the molecular result, so on resected nodules the sensitivity of both tests is overstated and the specificity of molecular testing is understated; after weighting, the DL model remained more specific than molecular testing (0.98 vs 0.71).

Our findings agree with the small body of prior work. Daniels \etal found a similarly specific but insensitive profile when predicting molecular risk from ultrasound (specificity 97\%, sensitivity 45\% on 53 test images) \citep{daniels2020machine}. Gild \etal reported an AUC of 0.75 in resected Bethesda III nodules \citep{gild2022risk}, and Keutgen \etal a cross-validated AUC of 0.67 \citep{keutgen2022machine}. Kim \etal reported the same AUC of 0.75 in Bethesda III nodules from five institutions \citep{kim2026artificial}. A combined ultrasound and molecular model reached an AUC of 0.83 versus 0.63 for molecular testing alone, although unoperated nodules were assumed to be benign \citep{athreya2025combining}. The list prices of ThyroSeq v3 and Afirma GSC have been reported at about \$4{,}000 and \$6{,}400 per test \citep{nicholson2019molecular}, whereas the DL model only uses ultrasound images that are already acquired for every nodule.

The study has several limitations, mainly related to the dataset itself. Even though we have 466 thyroid nodule records, which is not a small number, only 191 of the imaged records (188 unique nodules) have surgical pathology results, namely the ground truth. With so few labeled nodules, it is hard to substantially improve the deep learning algorithm, due to the data-hungry nature of CNNs. This also limited the testing dataset to 50 cases, reduced to 47 unique nodules after removal of duplicate records. With only nine malignant test nodules, the confidence intervals are wide, and the comparison of sensitivities in particular is underpowered. The study was conducted at a single institution, and the training and test sets differed in molecular platform (ThyroSeq v3 was used only in the test set), because record numbers were grouped by platform and the testing dataset was the last block of records; this limits the generalizability of the test-set comparison, and on all 188 labeled nodules analyzed together the two tests had similar AUCs. The nesting of DL-positive within molecular-positive nodules, and the separation of molecular-positive nodules by the DL output, were seen in the test set but not in the training set. All DL outputs come from a single prediction file generated during model development; we could not determine retrospectively whether it came from one run or was merged from several runs, or fully reconstruct the training history of the model version behind it. The test nodules were held out from all analyses in this paper, but because the DL outputs were generated before the testing dataset was assembled, we cannot confirm that these nodules were excluded from the data used to fine-tune the CNN. The values reported for the original model in our earlier analysis \citep{yang2024machine} were computed from the same prediction file as the results in this paper, on all 191 labeled records including the test nodules, so they do not measure the gain from fine-tuning; a paired comparison of the original and fine-tuned models on the same 47 test nodules is in progress. The operating points were inherited from the original algorithm rather than optimized for indeterminate nodules, the raw DL output was not calibrated to this population, and we did not compare the DL model with the ACR TI-RADS category assigned by the radiologist \citep{tessler2017acr}. Although all nodules were identified through an AUS/FLUS diagnosis, the cytologic category recorded at a second FNA differed in 37 of the 166 analysis nodules (Bethesda IV in 29, V in 3, II in 4 and I in 1), no AUS/FLUS cytology is recorded in our dataset for 8 nodules (2 in the test set), and which aspirate was sent for molecular testing was not recorded. We did not analyze the time between the ultrasound examination, the molecular test and surgery. Finally, the reference standard was available only for resected nodules; we quantified, but could not eliminate, the resulting verification bias. Because surgery after a negative molecular result also depended on nodule size (median 2.2 vs 1.9~cm in resected vs unresected nodules), on repeat cytology (Bethesda IV or V in 20\% vs 6\%) and on platform (58\% of negative 7-gene panel results were followed by surgery, against 13--14\% for Afirma GEC and ThyroSeq v2 and v3), the Begg--Greenes correction is approximate; stratified by platform, the corrected sensitivity of molecular testing was 0.53.

\section{Conclusion}
In a single-institution cohort of Bethesda III nodules, a CNN adapted to indeterminate nodules matched molecular testing in overall discrimination on held-out nodules (AUC 0.73 vs 0.55) and, at the TR5 cut-off, was considerably more specific, with a positive result increasing the odds of malignancy about eightfold. In the test set, all DL-positive nodules were molecular-positive, and the DL output identified a subgroup of molecular-positive nodules with a lower risk of malignancy, suggesting that ultrasound DL can complement molecular testing. The next step is to confirm these findings, and to evaluate DL-guided de-escalation prospectively, in larger, multi-institutional cohorts.

{\small
\bibliographystyle{ieeenat_fullname_etal}
\bibliography{references}
}

\clearpage
\appendix
\onecolumn
\section*{Supplementary Material}
\setcounter{table}{0}
\renewcommand{\thetable}{S\arabic{table}}

\begin{table}[h]
\centering
\caption{Referral strategies on the test set (n = 47, 9 malignant) combining the molecular result with the DL output at the TR3, TR4 and TR5 cut-offs. At the TR5 cut-off all DL-positive nodules were molecular-positive, so the ``and'' rule coincides with the DL model alone and the ``or'' rule with the molecular test alone.}
\label{tab:strategies}
\resizebox{\linewidth}{!}{\begin{tabular}{llccccc}
\toprule
\textbf{DL cut-off} & \textbf{Refer to surgery if} & \textbf{Operated} & \textbf{Benign operated} & \textbf{Cancers missed} & \textbf{Sensitivity} & \textbf{Specificity} \\
\midrule
TR3 & DL alone & 40 & 32 & 1 & 0.89 & 0.16 \\
TR3 & Molecular-positive and DL-positive & 30 & 24 & 3 & 0.67 & 0.37 \\
TR3 & Molecular-positive or DL-positive & 43 & 34 & 0 & 1.00 & 0.11 \\
\midrule
TR4 & DL alone & 34 & 26 & 1 & 0.89 & 0.32 \\
TR4 & Molecular-positive and DL-positive & 26 & 20 & 3 & 0.67 & 0.47 \\
TR4 & Molecular-positive or DL-positive & 41 & 32 & 0 & 1.00 & 0.16 \\
\midrule
-- & Molecular test alone & 33 & 26 & 2 & 0.78 & 0.32 \\
TR5 & DL alone & 6 & 2 & 5 & 0.44 & 0.95 \\
TR5 & Molecular-positive and DL-positive & 6 & 2 & 5 & 0.44 & 0.95 \\
TR5 & Molecular-positive or DL-positive & 33 & 26 & 2 & 0.78 & 0.32 \\
\bottomrule
\end{tabular}
}
\end{table}

\begin{table}[h]
\centering
\caption{Sensitivity analyses of sensitivity and specificity (95\% CIs). IPW: inverse-probability-of-verification weighting, with weights derived from the probability of a definitive surgical diagnosis given the molecular result among the 404 unique imaged nodules with a definitive molecular result that were in the analysis cohort or had no definitive surgical diagnosis. Begg--Greenes: correction of the molecular test in that population.}
\label{tab:robust}
\begin{tabular}{llcc}
\toprule
\textbf{Analysis} & \textbf{Index test} & \textbf{Sensitivity} & \textbf{Specificity} \\
\midrule
Primary (test set, n=47) & DL (TR5) & 0.44 (0.14--0.79) & 0.95 (0.82--0.99) \\
 & Molecular & 0.78 (0.40--0.97) & 0.32 (0.18--0.49) \\
\midrule
NIFTP counted as malignant (3 NIFTP) & DL (TR5) & 0.42 (0.15--0.72) & 0.97 (0.85--1.00) \\
 & Molecular & 0.83 (0.52--0.98) & 0.34 (0.19--0.52) \\
\midrule
Verification-weighted (IPW) & DL (TR5) & 0.23 (0.05--0.67) & 0.98 (0.94--1.00) \\
 & Molecular & 0.40 (0.13--1.00) & 0.71 (0.50--0.83) \\
Begg--Greenes (n=404) & Molecular & 0.45 (0.30--0.69) & 0.76 (0.71--0.81) \\
\midrule
Analysis cohort, Afirma (n=91) & Molecular & 1.00 (0.86--1.00) & 0.20 (0.11--0.31) \\
Analysis cohort, ThyroSeq (n=75) & Molecular & 0.56 (0.31--0.78) & 0.60 (0.46--0.72) \\
\quad Afirma vs ThyroSeq (Fisher) & & P $<$ 0.001 & P $<$ 0.001 \\
\bottomrule
\end{tabular}

\end{table}

\begin{table}[h]
\centering
\caption{Main comparison repeated on alternative evaluation sets. The originally assembled test set includes three duplicate records; ``all labeled nodules'' pools the training nodules, the test nodules and the 22 labeled nodules outside the analysis dataset (molecular results that were indeterminate are counted as negative, as in the original data files). Sensitivity, specificity and PPV refer to the TR5 cut-off; the last column gives the malignancy rate of molecular-positive nodules with a positive vs a negative DL result.}
\label{tab:sets}
\resizebox{\linewidth}{!}{\begin{tabular}{lcccccccc}
\toprule
\textbf{Evaluation set} & \textbf{n (malignant)} & \textbf{AUC DL} & \textbf{AUC Mol.} & \textbf{P} & \textbf{Sens. DL / Mol.} & \textbf{Spec. DL / Mol.} & \textbf{PPV DL / Mol.} & \textbf{Mol.+: risk DL+ vs DL$-$} \\
\midrule
Primary test set (deduplicated) & 47 (9) & 0.73 & 0.55 & 0.15 & 0.44 / 0.78 & 0.95 / 0.32 & 0.67 / 0.21 & 4/6 vs 3/27 (P = 0.01) \\
Test set as originally assembled & 50 (10) & 0.76 & 0.55 & 0.07 & 0.50 / 0.80 & 0.95 / 0.30 & 0.71 / 0.22 & 5/7 vs 3/29 (P = 0.003) \\
All labeled nodules in the database & 188 (50) & 0.59 & 0.60 & 0.78 & 0.34 / 0.78 & 0.78 / 0.43 & 0.35 / 0.33 & 13/31 vs 26/87 (P = 0.27) \\
\bottomrule
\end{tabular}
}
\end{table}

\end{document}